\documentclass[10pt,onecolumn]{article}

\usepackage{graphicx}
\usepackage{epstopdf}
\usepackage{dcolumn}
\usepackage{bm}

\usepackage{amssymb}
\usepackage{amsmath}
\usepackage{amsfonts}

\title{Thermodynamics of hot quantum scalar field in a  (D +1) dimensional curved spacetime}
\author{W. A. Rojas C.\thanks{e-mail: warojasc@unal.edu.co}\\ Departamento de F\'isica \\ Universidad Nacional de Colombia,  Bogot\'a, Colombia\vspace{0.2cm} \\
J. R. Arenas S.\thanks{e-mail: jrarenass@unal.edu.co}\\ Observatorio Astron\'omico Nacional\\ Universidad Nacional de Colombia, Bogot\'a, Colombia 
}

\begin{document}

\maketitle






\begin{abstract}
We use the brick wall model to calculate the free energy of  quantum scalar field in a curved spacetime (D +1) dimensions.
We find the thermodynamics properties of quantum scalar field in several scenaries: Minkowski spacetime, Schwarzschild spacetime and BTZ spacetime. For the cases analysed, the thermodynamical properties of quantum scalar field is exactly with the reported.
 It was found that the entropy of the gas is proportional to the horizon area in a gravity field strong,  which is consistent with the holographic principle.



\end{abstract}



\section{Introduction}
In the early 70's, a connection between gravity and thermodynamics is established. Where the horizon for a black hole has an associated temperature 
	\[T_{H}=\frac{\hbar \kappa_{0}}{2\pi}
\]
 and entropy 
	 \[S_{BH}=\frac{1}{4 l^{2}_{pl}}A.
 \]
This entropy is proportional to black hole surface. The Bekenstein-Hawking entropy is considered a true thermodynamic entropy of  the black holes. In a typical thermodynamic system, the thermal properties must reflect the microscopic physics. The temperature is a measure of average energy of particles and entropy counts the number of microstates of the system.

The first models emerged to explain the origin of entropy arising from Gibbons/Hawking studies and model t'Hooft brick wall. For the former model (Euclidean approach) provides no insight into the  dynamical  origin of $S_{BH}$. 

The latter (brick wall model) studies behavior of scalar fields near to black hole horizon \cite{MU}. Under the last model, the  $S_{BH}$ is related with the vacuum fluctuations in strong gravitational fields \cite{FU}. Because, an observer it at rest with black hole horizon sees a thermal bath of particles on the horizon \cite{MU}. 

In this point, we must clarify the concept of brick wall. Following to Israel \cite{IS}:  

\textit{The brick wall is treated here as a real physical barrier. To prevent misunderstanding this conception must be distinguished from a quite different one, according to which the wall is fictitious, merely a mathematical \textbf{cutoff} used to regularize the $(\Delta r)^{-1}$ and $log \Delta r$ terms in the expression for the thermal entropy} .

Our main purpose in this paper is study of hot quantum scalar field in a (D +1) dimensional curved spacetime. Under the  brick wall model, we calculate the free energy $F_{D}$ of  hot quantum scalar field  in (D+1) dimensions. 

For this study D are n dimensional space, where n=2,3... This method allows find the black hole entropy $S_{D}$ and this is proportional to area. 

In the section two,  we calculate the free energy for hot quantum scalar field in a (D+1) dimensional curved spacetime. Under standard prescription we derive the entropy for this scenery.

In the section three, we revised the thermodynamics of a hot quantum field in (3+1) dimensional flat spacetime. 

For the section three we study the particular case of D=3. We find that the thermodynamics properties are same reported for \cite{FU} and \cite{RO}.

In the section four, we study the BTZ black hole, we find that entropy under method explained en the section two is proportional al area.

In the end, we present  discussions, we conclude that this model corresponds to a generalization of Brick wall model in (D + 1) dimensions.

\section{Thermodynamics of ideal gas in (D+1) dimensions}

We consider a spacetime with metric
\begin{equation}
ds^{2}=-f(r)dt^{2}+\frac{1}{f(r)}dr^{2}+r^{2}d\Omega^{2}
\end{equation}
this includes the spacetimes Schwarzschild, Reissner-Nordstronm and (anti-) de Sitter in spacetime (3+1) dimensions \cite{MU}.

For this spacetime Fursaev showed that free energy  $F_{3}$ for a bosonic field in three spatial dimensions near  to black hole horizon is \cite{FU}
\begin{equation}
F_{3}=-\frac{1}{\pi^2}\zeta(4)\int T^{4} \sqrt{-g}d^{3}x,
\end{equation}
where   $\zeta(4)$ is Riemann function zeta, $T$ is temperature of the bosonic gas, $g$ is the  determinant of metric and $d^3x$ is differential element of volume spatial. 

Free energy in (2 +1) for a gas of bosons is 
	\[F_{2}= -\frac{1}{\pi}\zeta(3)\int T^{3} \sqrt{-g}d^{2}x.
\]
If we consider a (D+1)-dimensional spacetime with a metric \cite{KO}
\begin{equation}
ds^{2}=-f(r)dt^{2}+\frac{1}{f(r)}dr^{2}+r^{D-1}d\Omega^{D-1}.
\end{equation}
and metric tensor $g_{\mu\nu}$ giveb by
\begin{equation}
g_{\mu \nu}= \left(
	\begin{array}{ccc} -f(r)& \ldots& 0 \\ \vdots &\frac{1}{f(r)} &\ldots \\ 0 &\ldots& r^{D-1}\Omega
	\end{array}
	\right),
\end{equation}
Then, the free energy is
\begin{equation}
F_{D}=-\frac{1}{\pi^{D-1}}\zeta(D+1)\int T^{D+1} \sqrt{-g}d^{D}x,
\end{equation}
where $T$ is temperature of ideal gas and obey the Tolman's law
\begin{equation}
T(r)=\frac{T_{\infty}}{f(r)^{1/2}}
\end{equation}
and $g$ is the determinant of metric tensor
\begin{equation}
g=r^{D-1}\Omega.
\end{equation}
Their square root is 
\begin{equation}
\sqrt{-g}=\sqrt{r^{D-1}\Omega}=r^{\frac{D-1}{2}}\Omega^{1/2},
\end{equation}
where $r$ is radial part and $\Omega$ is solid angle in $D$ dimensions. The differential element of volume spatial $d^{D}x$,  is the product of a radial part ($dr$) and an angular part ($d^{D-1}\Omega$)
\begin{equation}
d^{D}x=dr d^{D-1}\Omega.
\end{equation}
Thus, the free energy is reduced to
\medskip{}
\begin{equation}\label{FE1}
F_{D}=-\frac{1}{\pi^{D-1}} \zeta(D+1) \int \frac{T^{D+1}_{\infty}}{f(r)^{\frac{D+1}{2}}}  r^{\frac{D-1}{2}}dr  \Omega^{1/2} d^{D-1}\Omega.
\end{equation}
Separating radial part and angular part in  (10), we can write free energy as
\begin{equation}
F_{D}=-\frac{1}{\pi^{D-1}}\zeta(D+1) T^{D+1}_{\infty}  \int \frac{r^{ \frac{ D-1}{2}}}{f(r)^{\frac{D+1}{2}}}dr   \int \Omega^{1/2}d^{D-1}\Omega.
\end{equation}
The first integral in (11) is associated to radial part and second one to angular part. Following S. Kolelar and  T. Padmanabhan \cite{KO}, (11) can be written as
\begin{equation}
F_{D}=-\frac{1}{\pi^{D-1}}\zeta(D+1) T^{D+1}_{\infty}  \Omega \int\frac{r^{\frac{D-1}{2}}}{f(r)^{\frac{D+1}{2}}}dr.
\end{equation}
Where $\Omega$ is solid angle in $D$ dimensions.

Consider the  following conditions \cite{KO}: 
	\[f(r_{H})=0,
\]
which defines the horizon 
	\[2\kappa=f'(r_{H})
\]
It is a finite temperature in the vicinity of the black hole. Near to horizon, we write $f(r)$ as \cite{MU,KO,SU}
\begin{equation}
f(r)=f'(r_{H})(r-r_{H}).
\end{equation}
According to the above, the integral of (12) reduces to
\begin{equation}
	\int\frac{r^{\frac{D-1}{2}}}{f(r)^{\frac{D+1}{2}}}dr=\int\frac{r^{\frac{D-1}{2}}}{\left(f'(r_{H})(r-r_{H})\right)^{\frac{D+1}{2}}}dr
=\frac{1}{f'(r_{H}) ^{\frac{D+1}{2}}}\int\frac{r^{\frac{D-1}{2}}}{(r-r_{H})^{\frac{D+1}{2}}}dr.
\end{equation}
Defining $u=r-r_{H}$ 
\begin{equation}
\frac{1}{f'(r_{H}) ^{\frac{D+1}{2}}}\int\frac{r^{\frac{D-1}{2}}}{(r-r_{H})^{\frac{D+1}{2}}}dr=\frac{1}{f'(r_{H}) ^{\frac{D+1}{2}}}\int \frac{(u+r_{H})^{\frac{D-1}{2}}}{u^{\frac{D+1}{2}}}du.
\end{equation}

And using a binomial expansion
\begin{equation}
(r_{H}+u)^{\frac{D-1}{2}}=r^{\frac{D-1}{2}}_{H}+\frac{1}{1!}\left(\frac{D-1}{2}\right)r^{\frac{D-3}{2}}_{H}u+\frac{1}{2!}\left(\frac{D-1}{2}\right)\left(\frac{D-3}{2}\right)r^{\frac{D-3}{2}}_{H}u^2+\ldots+u^{\frac{D-1}{2}}.
\end{equation}
So, (15) rewrites
	\[\frac{1}{f'(r_{H}) ^{\frac{D+1}{2}}}\int \frac{(u+r_{H})^{\frac{D-1}{2}}}{u^{\frac{D+1}{2}}}du
\]
\begin{equation}
=\frac{1}{f'(r_{H}) ^{\frac{D+1}{2}}}\int \frac{1}{u^{\frac{D+1}{2}}}           
\left[r^{\frac{D-1}{2}}_{H}+\frac{1}{1!}\left(\frac{D-1}{2}\right)r^{\frac{D-3}{2}}_{H}u+\frac{1}{2!}\left(\frac{D-1}{2}\right)\left(\frac{D-3}{2}\right)r^{\frac{D-3}{2}}_{H}u^2+\ldots+u^{\frac{D-1}{2}} \right]   du.
\end{equation}

Then the integral in (15) is rewritten as (17) \cite{KO}.

S. Kolelar and  T. Padmanabhan says: the main contribution to this integral comes from the lower limit of the integral $\frac{r^{\frac{D-1}{2}}_{H}}{u^{\frac{D+1}{2}}}$.  Because near the horizon $u=h$  and $\frac{h}{r_{H}}\ll l_{P}$ \cite{KO}.
\begin{equation}
\frac{1}{f'(r_{H}) ^{\frac{D+1}{2}}}\int \frac{r_{H}^{\frac{D-1}{2}}}{u^{\frac{D+1}{2}}}du =-\frac{2 r^{\frac{D-1}{2}}_{H}}{f'(r)^{\frac{D+1}{2}}(D-1)h^{\frac{D-1}{2}}}
\end{equation}
where $l_{p}$ is the Planck length. Near the horizon, we can replace $h$ to $l_{p}$ as a distance own
	\begin{equation}
	l_{p}=\int^{H}_{r_{H}}\frac{dr}{\sqrt{f(r)}}\approx 2\sqrt{\frac{h}{f'(r_{h})}},
\end{equation}

solvig $h$  and replace in (18)
\begin{equation}
\int \frac{r^{\frac{D-1}{1}}}{f(r)^{\frac{D+1}{2}}}dr=-\frac{2^{D}r_{H}^{\frac{D-1}{2}}}{(D-1)\left[f'(r_{H})\right]^{D}l_{p}^{D-1}}
\end{equation}

Finally we write the free energy for ideal gas in (D+1) dimensions in the following form

\begin{equation}
F_{D}=-\frac{\zeta(D+1)}{(D-1)\pi^{D-1}}\left[\frac{A_{\frac{D-1}{2}}}{l_{p}^{D-1}}\right]\frac{T_{\infty}^{D+1}}{\kappa^{D}}
\end{equation}

where the product $A_{\frac{D-1}{2}}=\Omega r_{H}^{{\frac{D-1}{2}}}$ is the horizon area in (D +1) dimensions. The free energy is an \textit{off-shell} prescription and expresses in four independent variables:
\begin{itemize}
	\item The temperature $T_{\infty}$,
	\item geometrical characteristics $(D+1)$ spacetime dimensions, horizon's area $A$ and surface gravity $\kappa$.	
\end{itemize}

The entropy is recoverable from the free energy by the standard prescription
\begin{equation}
S_{D}=-\left(\frac{\partial F}{\partial T}\right)
\end{equation}
\begin{equation}
S_{D}=	\frac{\zeta(D+1)}{(D-1)\pi^{D-1}}\left[\frac{A_{\frac{D-1}{2}}}{l_{P}^{D-1}}\right]\frac{(D+1)T_{\infty}^{D}}{\kappa^{D}}.
\end{equation}
At this point, we observe that this is an \textit{off-shell} prescription \cite{MU, FR}. Because, geometrical quantities (A y $\kappa$) are kept fixed when the temperature is varied \cite{MU}.
\section{Particular case (3+1) dimensional flat spacetime}
If we take the equation (5) and we do $g_{\mu \nu}=\eta_{\mu\nu}$ where $\eta_{\mu\nu}$ is  tensor minkowski for a flat spacetime. So the equation (5) become in
\begin{equation}
F_{D}=-\frac{1}{\pi^{D-1}}\zeta(D+1)\int{T^{D+1}\sqrt{-\eta}d^{D}x},
\end{equation}
where $\sqrt{-\eta}=1$, $T=T_{\infty}$ and $V_{D}=\int{d^{D}x}$. Then, we have
\begin{equation}
F_{D}=-\frac{1}{\pi^{D-1}}\zeta(D+1)T^{D+1}_{\infty}V_{D}.
\end{equation}
We  find the entropy of the hot quantum scalar field in the  Minkowski spacetime like
	\[S_{D}=-\frac{\partial F}{\partial T_{\infty}}
\]
\begin{equation}
S_{D}=\frac{(D+1)}{\pi^{D-1}}\zeta(D+1)T^{D}_{\infty}V_{D}.
\end{equation}
In particular  the case when  $D=3$ for an flat spacetime. The free energy is
\begin{equation}
F_{3}=-\frac{\pi^{2}}{90}T^{4}_{\infty}V_{3}
\end{equation}
and the  entropy is
\begin{equation}
S_{3}=\frac{2}{45}\pi^{2}T^{3}_{\infty}V_{3}
\end{equation}
other properties of hot quantum scalar field in the minkowski spacetime are well known \cite{LA}.
\section{Particular case $(3+1)$ \\ dimensional curved spacetime}

If the temperature of quantum field is $T_{\infty}=T_{H}=\frac{\kappa}{2 \pi}$, so the entropy is $S_{3}|_{T_{\infty}=T_{H}}$ using (23)
	\[S_{3}|_{T_{\infty}=T_{H}}=\frac{1}{360 \pi l^{2}_{p}}A
\]
it is agree with Fursaev's result!! The  quantum field is  supposed to be in thermal equilibrium with the  horizon \cite{FR}.

Another hand, if D = 3, (21) and (23)  reduced to

\begin{equation}
F_{3}=-\frac{\pi^{2}}{180}\frac{T^{4}_{\infty}A}{l_{p}^{2}\kappa^{3}}
\end{equation}

\begin{equation}
S_{3}=\frac{\pi^{2}}{45}\frac{T^{3}_{\infty}A}{l_{p}^{2}\kappa^{3}}
\end{equation}

such that (26) is entropy of bosonic field in (3+1). For a photon gas in Schwarzschild spacetime, their  entropy is \cite{RO}
\begin{equation}
S_{3}=\frac{1}{45}\frac{\pi^{2}k^{4}_{B}c^{3}}{\hbar^{3}}\left[\frac{A}{l^{2}_{P}\kappa^{3}}\right] T^{3}_{\infty}.
\end{equation}

Also we calculate other thermodynamics properties with standard recipes:
\begin{itemize}
	\item The internal energy of quantum field is 
	\begin{equation}
	E_{D}=\frac{\zeta(D+1)}{(D-1)\pi^{D-1}}\left[\frac{A_{\frac{D-1}{2}}}{l^{D-1}_{p}}\right]\frac{D T^{D+1}_{\infty}}{\kappa^{D}}.
	\end{equation}
	If D=3 the internal energy 
		\begin{equation}
		E_{3}=\frac{\pi^{2}}{60}\frac{A T^{4}_{\infty}}{l^{2}_{p}\kappa^{3}},
		\end{equation}
		that  is the  internal energy of photon gas in Schwarzschild spacetime \cite{RO}.
		
	\item The specific heat is	
		\[
		C_{V}=\left(\frac{\partial E}{\partial T_{\infty}}\right)_{V}
	\]
	\begin{equation}
	C_{V}=\frac{\zeta(D+1)}{(D-1)\pi^{D-1}}\left[\frac{A_{\frac{D-1}{2}}}{l^{D-1}_{p}}\right]\frac{D(D+1)T^{D}_{\infty}}{\kappa^{D}},
	\end{equation}
	with $D=3$
		\begin{equation}
	C_{V-3}=\frac{\pi^{2}}{15}\frac{A T^{3}_{\infty}}{l^{2}_{p}\kappa^{3}}.
	\end{equation}	
	
\item And the pressure of quantum field near to horizon is
	\[P=-\left(\frac{\partial F}{\partial V}\right)_{T_{\infty}}=\frac{1}{l_{P}} \left(\frac{\partial F}{\partial A}\right)_{T_{\infty}}
	\]
		\begin{equation}
		P_{D}=\frac{\zeta(D+1)}{(D-1)\pi^{D-1}}\left[\frac{1}{l^{D}_{p}}\right]\frac{T^{D+1}_{\infty}}{\kappa^{D}}
			\end{equation}
		with $D=3$
		\begin{equation}
	P_{3}=\frac{\pi^{2}}{180}\frac{T^{4}_{\infty}}{l^{3}_{p}\kappa^{3}}.
	\end{equation}	
\end{itemize}
Is interesting to note that the thermodynamic properties described by (24), (26), (28), (30) and (32) have already been studied by the authors in the case of a photon gas in the Schwarzschild spacetime \cite{RO} under the approximation of brick wall model by Mukohyama and Israel for a hot quantum scalar field near the horizon in the Boulware state \cite{MU}.


\section{BTZ spacetime}

This spacetime is interesting  because the BTZ black holes are asymptotically anti de Sitter. In gravity (2+1) the curvature is constant ($R=-6/l^{2}$). This black holes don't have points and regions in which the curvature is divergent. However, the BTZ black holes have a horizon, an ergosphere and thermodynamic properties to the  classical solutions of General Relativity \cite{LAR}. 


Their metric is
\begin{equation}
ds^{2}=-\left(\frac{r^{2}-r^{2}_{+}}{l^{2}}\right)dt^{2}+\frac{1}{\left(\frac{r^{2}-r^{2}_{+}}{l^{2}}\right)}dr^{2}+r^{2}d\phi,
\end{equation}
with a negative cosmological constant $\lambda=-\frac{1}{l^{2}}$.
The area of horizon of BTZ black hole is the length $2\pi r_{+}$ \cite{FU, BA}.

We calculate the black hole entropy for this spacetime with method of the section II. The first step is to use  (5)  in which D=2 spatial dimensions, then we have
\begin{equation}
F_{2}=-\frac{1}{\pi}\zeta(3)\int T^{3}\sqrt{-g}d^{2}x,
\end{equation}
in agree with (6), (8) $\sqrt{-g}=r^{1/2}\Omega^{1/2}$ and (9) $d^{2}x=dr d\Omega$. The free energy in the BTZ spacetime is rewritten as
\begin{equation}
F_{2}=-\frac{1}{\pi}\zeta(3) \int \left[\frac{T_{\infty} }{f(r)^{1/2}}\right]^{3}r^{1/2}\Omega^{1/2} dr d\Omega=-\frac{1}{\pi}\zeta(3)T^{3}_{\infty}\int{\Omega^{1/2}d\Omega}\int\frac{r^{1/2}}{f(r)^{3/2}}dr,
\end{equation}
where the solid angle is $\Omega=\int \Omega^{1/2}d\Omega=2\pi$ and $f(r)=\left(\frac{r^{2}-r^{2}_{+}}{l^{2}}\right)$ for the  BTZ spacetime.
We consider the last integral is agree with the condition (13), the  integral reduces to
\begin{equation}
\int\frac{r^{1/2}}{f(r)^{3/2}}dr=\frac{1}{\left[f'(r_{+})\right]^{3/2}}\int\frac{r^{1/2}}{(r-r_{+})^{3/2}}dr,
\end{equation}
this integral is similar to (14) y it can solve the same form. With $u=r-r_{+}$, thus (41) become in
\begin{equation}
\int\frac{r^{1/2}}{f(r)^{3/2}}dr=\frac{1}{\left[f'(r_{+})\right]^{3/2}}\int\frac{(r_{+}+u)^{1/2}}{u^{3/2}}du.
\end{equation}
And using a binomial expansion
\begin{equation}
(r_{+}+u)^{1/2}=r_{+}^{1/2}+\frac{1}{2}r_{+}^{-1/2}u+ \ldots +u^{1/2}.
\end{equation}
Then the integral (42) is rewritten as
	\[\frac{1}{\left[f'(r_{+})\right]^{3/2}}\int\frac{(r_{+}+u)^{1/2}}{u^{3/2}}du=
\]
\begin{equation}
=\frac{1}{\left[f'(r_{+})\right]^{3/2}}\int\frac{\left[r_{+}^{1/2}+\frac{1}{2}r_{+}^{-1/2}u+ \ldots +u^{1/2}\right]}{u^{3/2}}du\approx\frac{1}{\left[f'(r_{+})^{3/2}\right]}  \int \frac{r_{+}^{1/2}}{u^{3/2}}du=-\frac{2r^{1/2}_{+}}{f'(r_{+})^{3/2}}.\frac{1}{\sqrt{u}}
\end{equation}
under the condition of S. Kolekar and  T. Padmanabhan \cite{KO}. Again, near to horizon $u=h$ and we replace $h$ to $l_{p}$ as distance own in agree with (19)
\begin{equation}
\int \frac{r^{1/2}}{f(r)^{3/2}}=-\frac{2^{2}r_{+}^{1/2}}{\left[f'(r)\right]^{2}l_{p}}.
\end{equation}

We write  the free energy for hot quantum scalar field in BTZ spacetime as
\begin{equation}
F_{2}=-\frac{1}{\pi}\zeta(3)T^{3}_{\infty}\left[\Omega r^{1/2}_{+}\right]\frac{2^2}{2^2\kappa^{2}l_{p}}=-\frac{1}{\pi}\zeta(3)\frac{T^{3}_{\infty}}{\kappa^{2}}\frac{\left[A_{\frac{2-1}{2}}\right]}{l_{p}}
\end{equation}
where the area of horizon is $A_{\frac{2-1}{2}}=\Omega r^{1/2}_{+}=2\pi r_{+}$. We can obtain  the entropy as
\begin{equation}
S_{2}=\frac{3}{\pi}\zeta(3)\frac{T^{2}_{\infty}}{\kappa^{2}}\left[ \frac{2\pi r_{+}}{l_{p}}\right].
\end{equation}

Again, if the  temperature of quantum field is $T_{\infty}=T_{H}=\frac{\kappa}{2\pi}$  and the entropy to the temperatura Hawking is
	\[
	S_{2}|_{T_{\infty}=T_{H}}=\frac{3}{\pi} \zeta(3) \frac{1}{\kappa^{2}}\frac{\kappa^{2}}{4\pi^{2}}\left[ \frac{2\pi r_{+}}{l_{p}}\right]
\]
\begin{equation}
S_{2}|_{T_{\infty}=T_{H}}=\frac{2\pi r_{+}a}{l_{p}}.
\end{equation}
The mass $M$ of the BTZ black hole is defined as
\begin{equation}
M=\frac{r^{2}_{+}}{8l^{2}G_{3}},
\end{equation}
where $G_{3}$ is the 3D gravitational coupling and has the dimension of length \cite{FU}.

Thus, $S_{2}|_{T_{\infty}=T_{H}}$ is
 \begin{equation}
S_{2}|_{T_{\infty}=T_{H}}=\frac{2\pi a}{l_{p}}\sqrt{8l^{2}MG_{3}}
\end{equation}

\section{Summary and Discussion}

The last result shows that entropy of ideal gas is proportional to area in (D+1) dimensions, when this is in  equilibrium state with the  horizon at  temperature $T_{\infty}$.

Under the  brick wall model for (D+1) dimensional  curved spacetime, the  entropy can be determined by response of the  free energy of the system to change of temperature given by (22) \cite{FU, MU, RO,  FR}. Also the  distinction between \textit{thermodynamical} and \textit{statistical} entropies disappears in this model, because the geometrical and thermal variables are kept independent. Observe that this is an “off-shell” prescription \cite{ MU, FR}.

 The free energy was made from the brick wall model, which studies quantum fields close to the horizon \cite{MU}. Following this model, thus affirms Fursaev  and Mukohyama:  entropy defined in (23) and corresponds to the Bekenstein-Hawking entropy $S_{BH}$ for quantum fields near the horizon in spacetime (D +1). And also depends on the geometrical and thermal quantities as expressed Mukohyama \cite{FU, MU} 
\begin{equation}
S_{D} \propto \left[\frac{A_{\frac{D-1}{2}}}{(D-1)l_{P}^{D-1}}\right]\frac{(D+1)T_{\infty}^{D}}{\kappa^{D}}
\end{equation}
if $T=T_{H}$, then $S_{D}=S_{BH}$, that are defined only for  geometrical quantities in  (D+1) dimensions \cite{MU}. 

We find that the entropy to  the Hawking temperature in BTZ black hole is  proportional to
 \begin{equation}
S_{2}|_{T_{\infty}=T_{H}} \propto  2\pi \sqrt{l^{2}M},
\end{equation}
This result is consistent with \cite{ FU, BA}.

From the above we can consider that the present model is a generalization of the study of hot quantum scalar field  in a (D +1) dimensional curved spacetime  of brick wall model. In the case when $D = 3$ the usual space, the entropy is reduced to the results reported by \cite{FU} and \cite{RO}.


\section*{Acknowledgements}
\noindent

This work was supported by  the Departamento de Administrativo de Ciencia, Tecnolog\'ia e Innovaci\'on, Colciencias.


\end{document}